\documentclass[amsmath,showkeys,superscriptaddress,onecolumn,12pt]{revtex4-2}
\usepackage{float}
\usepackage[left=0.6in, right=0.6in, top=0.7in, bottom=0.7in]{geometry}
\usepackage{amsmath}
\usepackage{amsmath}
\usepackage{amssymb}
\usepackage{graphicx}
\usepackage[colorlinks=true,linkcolor=blue,citecolor=blue,urlcolor=blue,breaklinks]{hyperref}
 
\usepackage[colorlinks=true,linkcolor=blue,citecolor=blue,urlcolor=blue,breaklinks]{hyperref}

\usepackage{subfig}

\usepackage{float}
\newcommand{\subfigimg}[3][,]{%
	\setbox1=\hbox{\includegraphics[#1]{#3}}% Store image in box
	\leavevmode\rlap{\usebox1}% Print image
	\rlap{\hspace*{1.51cm}\raisebox{\dimexpr\ht1-1.1\baselineskip}{#2}}% Print label
	\phantom{\usebox1}% Insert appropriate spcing
}
\usepackage[utf8x]{inputenc}

\begin{document}
\title{Multipartite entanglement and purity dynamics in classical channels influenced by fractional Gaussian noise} 
\author{Muhammad Javed}
\address{Quantum Optics and Quantum Information Research Group, Department of Physics, University of Malakand, Chakdara Dir, Pakistan}
\author{Atta Ur Rahman}
\email{a.rahman@whu.edu.cn}
\address{Key Laboratory of Aerospace Information Security and Trusted Computing, Ministry of Education, School of Cyber Science and Engineering, Wuhan University, China}
\author{Lionel Tenemeza Kenfack}
\address{Department of Physics, Research Unit of Condensed Matter, Electronic and Signal Processing, University of Dschang, PO Box: 67, Dschang, Cameroon}
\author{Salman Khan Safi}
\address{COMSATS University, Park Road, Chak Shahzad, Islamabad, PO Box: 44000, Pakistan}

%
%\address{
%$^{1}$ Quantum Optics and Quantum Information Research Group, Department of Physics, University of Malakand, Chakdara Dir, Pakistan\\
%$^{2}$ Key Laboratory of Aerospace Information Security and Trusted Computing, Ministry of Education, School
%of Cyber Science and Engineering, Wuhan University, P.O. Box 430072, Wuhan, China\\
%$^{3}$ Department of Physics, Research Unit of Condensed Matter, Electronic and Signal Processing, University of Dschang, PO Box: 67, Dschang, Cameroon\\
%$^{4}$ COMSATS University, Park Road, Chak Shahzad, Islamabad, PO Box: 44000, Pakistan
%}

\begin{abstract}
The dynamical map of entanglement and mixedness in four-qubit maximally entangled GHZ state paired with classical channels driven by fractional Gaussian noise is investigated. The qubit-channel coupling is assumed in four distinct ways: common, bipartite, tripartite, and independent local channel-qubit configurations comprising single, double, triple, or independent noisy sources. Using entanglement witness, negativity, purity and von Neumann entropy, except for the independent configuration, we show that indefinite entanglement and purity preservation may be simulated in multipartite GHZ-like states. Quantum correlations and purity decrease exponentially in four qubits, and exact fluctuating local field behaviour, as well as entanglement sudden death and birth revivals, are completely suppressed. Entanglement and purity preservation are affected by noise and the number of independent channels utilised. The Hurst parameter of fractional Gaussian noise was discovered in the four qubits to improve entanglement and avoid mixedness. Different entanglement and mixedness criteria yield different results, emphasizing the need to investigate various approaches in order to uncover multipartite correlations.
\end{abstract}
%\noindent{\bf Keywords}: entanglement, mixedness, classical channels, fractional Gaussian noise
\maketitle
\section{Introduction}
Quantum computers have emerged as leading-edge advanced devices that are far superior in functionality and most certain in practical applications. The quantum mechanical phenomena which control the tasks to be accomplished are the major preoccupations and working concepts of quantum computers. Therefore, physical phenomena and the corresponding functional concepts are just as important to consider as quantum computers \cite{11,12}. Entanglement coherence and purity, among many other phenomena, are undeniably the most significant for efficient quantum mechanical operations \cite{13, 15}. Without a doubt, entanglement coherence and purity are at the root of nearly all quantum computing applications. For this reason, the dynamics of entangled and coherent quantum systems remain a valuable pathway. As a result, these phenomena are discovered to be extremely resourceful and to play a vital role in the successful practical deployment and better functioning of quantum mechanical protocols.
\par

Because of the non-local nature that classical interpretations cannot account for, entanglement is often used to distinguish quantum and classical correlations \cite{18,19}. Thus, entanglement has received a lot of attention and is being studied for new applications, such as quantum communication \cite{20}, quantum cryptography \cite{21}, dense coding \cite{22}, teleportation \cite{24}, quantum secure direct communication and quantum key distribution \cite{25,26}. For effective quantum information operations, we need entanglement preservation in quantum processing to execute various quantum information processing protocols \cite{29}. Besides, coherence also is at the forefront of quantum computation, where the "0" and "1" states of the qubits superpose, which leads to an acceleration in many traditional algorithms. But when a state is decoherent, all its quantumness is lost and its usefulness drops.

\par
The quantum information in progress is frequently connected to a surrounded medium and can never be fully isolated \cite{31,32,33}. One of the key drawbacks of the effective use of quantum information processing protocols is the physical interaction between quantum systems and their environments that results in non-local correlation's decline degradation and are termed as disentanglement and mixedness \cite{34,35,36}. The environmental defects that are responsible for these dephasing effects are known as environmental noise, which limits the system's initial state entanglement and purity of quantum systems. The environmental noise has two different interpretable forms, which are classical and quantum interaction pictures. We will use the classical noisy interpretation instead of the quantum one because it allows us to explore the time evolution of the quantum systems with a larger number of degrees of freedom \cite{37,38,39,40,41,42,43}. For practical quantum information processing, it is critical to optimize and describe the evolution of quantum systems in the presence of such fatal interfering classical environmental noises in particular. This can provide solutions to avoid interruption and losses in the quantum mechanical protocols.
\par

Several quantum systems have been extensively researched to date for the dynamics and protection of quantum correlations and purity under different classical environmental noisy effects. This includes static, dynamic, and coloured noises, all of which have yielded significant outcomes. Quantum correlations were increasingly fragile in the presence of these noises, and they were deteriorated or entirely disappeared \cite{34,35,36,37,38,39,40,41,42,43}. In addition, recent quantum research advances explore important measures, methods, and dynamical characterizations of different entangled quantum systems, such as bipartite, tripartite, and multi-party qubit systems. Here, multi-qubit systems outperform single-qubit and two-qubit systems in terms of channel capacities, cryptographic behaviour, and information transmission \cite{44,45,46,47}. We will study the dynamics of a multi-partite system because of the advantages it has over other quantum systems.
\par

This paper investigates the dynamics of the entanglement and purity for a system of four non-interacting qubits initially prepared as maximally entangled Greenberger-Horne-Zeilinger (GHZ) state when coupled to the classical noisy channels described by fractional Gaussian (FG) noise \cite{48,49,50}. This kind of Gaussian noise arises due to the random motion of the particles in the diffusion process. In the current study, the dynamics of the quantum systems can also undergo such disorder hence, should be carefully analyzed. The FG noise has been used to investigate weather information, traffic control analysis, and electrical measurements because of its higher time scale correlations \cite{KAPLAN}. This noise has also been used to obtain Tsallis permutation entropy \cite{plastino}, long-range and short-range dependence \cite{Murad} and multi-scale pattern images \cite{pentland}.

Here, we suggest coupling the four non-interacting qubits to the classical fluctuating environments in four different configurations: common local channel-qubit (CLCQ), bipartite local channel-qubit (BLCQ), tripartite local channel-qubit (TLCQ), and independent local channel-qubit (ILCQ) configurations. We coupled all four qubits to a common channel in the first case. In the second case, two pairs of qubits are independently coupled to two classical channels. While in the third situation, the state is coupled with three channels and, each qubit in the ILCQ configuration is connected with four individual classical channels. The focus of the current configurations is to study the variation in the FG noisy effects over the initially encoded non-local correlations and purity in a system of four qubits. In the presence of the current noisy situation, various measures such as quantum negativity, purity, and decoherence will evaluate the dynamics and protection of the multi-partite entanglement, purity and mixedness in the four qubits.

This paper is assembled as: In Sec. \ref{$QC$ measures}, the multipartite entanglement, purity, and mixedness estimators are illustrated. Sec. \ref{The Model} describes the physical model, FG noise application, and various types of system-channel configurations. Sec. \ref{Results and Discussions}, deals with the findings and the subsequent discussions. Sec. \ref{Conclusion}, represents the conclusive remarks based on the investigation undertaken.
\section{Entanglement and coherence measures} \label{$QC$ measures}
This section describes the quantifiers used to evaluate multipartite entanglement and coherence.

% ###################################################   NEW SUBSECTION #######################################################################

\subsection{Entanglement witness}\label{dd}
An entanglement witness is a function in quantum information theory that distinguishes a particular entangled state from separable ones. Both theoretically and experimentally, the entanglement witness operation has been successfully used to examine bipartite, tripartite, and multipartite entanglement in different quantum systems \cite{EW1,EW2,EW3}. Entanglement witness operation for a time evolved density matrix $\rho_{abcd}(t)$ can be computed as \cite{52}:
\begin{equation}
\text{ER}=-\hbox{Tr}[(\frac{1}{2}\textit{I}-\rho_o) \rho_{abcd}(t)].\label{ewo}
\end{equation}
Where $\rho_o$ is the initial density matrix for the current four-qubit GHZ state. For $\text{ER}=0$, the state will be separable, while the positive outcomes of the entanglement witness will imply a strict entanglement regime. Besides that, negative results may refer to both separability and a weak entanglement regime.

\subsection{N-partite negativity}\label{dd1}
Entanglement in an arbitrary N-qubit entangled system in a mixed state can be quantified by taking the average of the bipartite entanglement measures over all the possible bipartitions of the N-qubit system \cite{ltk1}. This is among the most practical strategies proposed for the quantification of the global amount of genuine entanglement in multipartite systems and can be mathematically expressed as:
\begin{equation}
\mathcal{N}^{(N)}(\rho)=\dfrac{2}{N}\sum\limits_{k=1}^{N/2}\left( \dfrac{1}{n_{bipart}^{(k)}}\sum\limits_{P=1}^{n_{bipart}^{(k)}}\mathcal{N}^{P[k|N-k]}(\rho)\right),\label{negativity}
\end{equation}
where $ k\arrowvert N-k $ represents the bi-partitions of the N-qubit system with $ k $ qubits in one block and the remaining $ N-k $ ones in another block. $ P\left[ k\arrowvert N-k\right]  $ is used to specify a precise combination of $ k $ and $ N-k $ qubits in constituting the bipartition $ k\arrowvert N-k $. Thus, $ n_{bipart}^{(k)} $ stands for the total possible non-equivalent concrete bi-partitions $ P\left[ k\arrowvert N-k\right] $. $ \mathcal{N}^{P[k|N-k]}(\rho) $ denotes the bipartite entanglement (in terms of negativity) associated to the concrete bipartition $ P\left[ k\arrowvert N-k\right] $. As mentioned above, the entanglement associated with any bipartition of the N-qubit system is evaluated by means of negativity, defined for an arbitrary N-qubit system in a mixed state $ \rho $ as:
\begin{equation}
\text{NY}=\sum\limits_{\jmath}\lvert\lambda_{\jmath}(\rho^{T_{I}})\rvert-1.
\end{equation}
Where $ \lambda_{\jmath}(\rho^{T_{I}}) $ are the eigenvalues of the partial transpose $ \rho^{T_{I}} $ of the total density matrix regarding the subsystem $ I $ which is constituted by the $ k $ qubits of the given bipartition $ P\left[ k\arrowvert N-k\right] $.
%###########################################################################################################################################
\subsection{Purity}
Purity is the quantifier of pureness and coherence in a quantum system. This measure mathematically can be written as \cite{53}:
\begin{equation}
\text{PY}=\hbox{Tr}[\rho_{abcd}(t)^2]. \label{purity}
\end{equation}
Where ${\rho_{abcd}(t)}$ is the time evolved density matrix of the system and for $m$-dimensional system, the multipartite state will be completely disordered and decoherent at $P(t)=\frac{1}{m}$. For a completely pure and coherent state, $P(t)=1$. Any other values between this upper and lower bound will show the corresponding amount of purity and coherence.
\subsection{Von Neumann entropy}
The quantum information decay caused by the unavoidable interaction of the environment with the qubits, which causes them to change their quantum states uncontrollably, is referred to as mixedness and decoherence. Von Neumann entropy approach here will estimate the mixedness in the current four-qubit state and can be computed by \cite{53}
\begin{equation}
V(t)=-\hbox{Tr}[\rho_{abcd}(t)ln\rho_{abcd}(t)],\label{decoherence}
\end{equation}
where $\rho_{abcd}(t)$ is the time evolved density matrix of the system. With the coherent quantum condition, $V(t)=0$ while, any other output value of this measure will display the corresponding mixedness amount.
\section{The model and dynamics}\label{The Model}
\begin{figure}[ht]

		\begin{tabular}{@{}p{0.45\linewidth}@{\quad}p{0.45\linewidth}@{}}
			\subfigimg[width=\linewidth]{(a)}{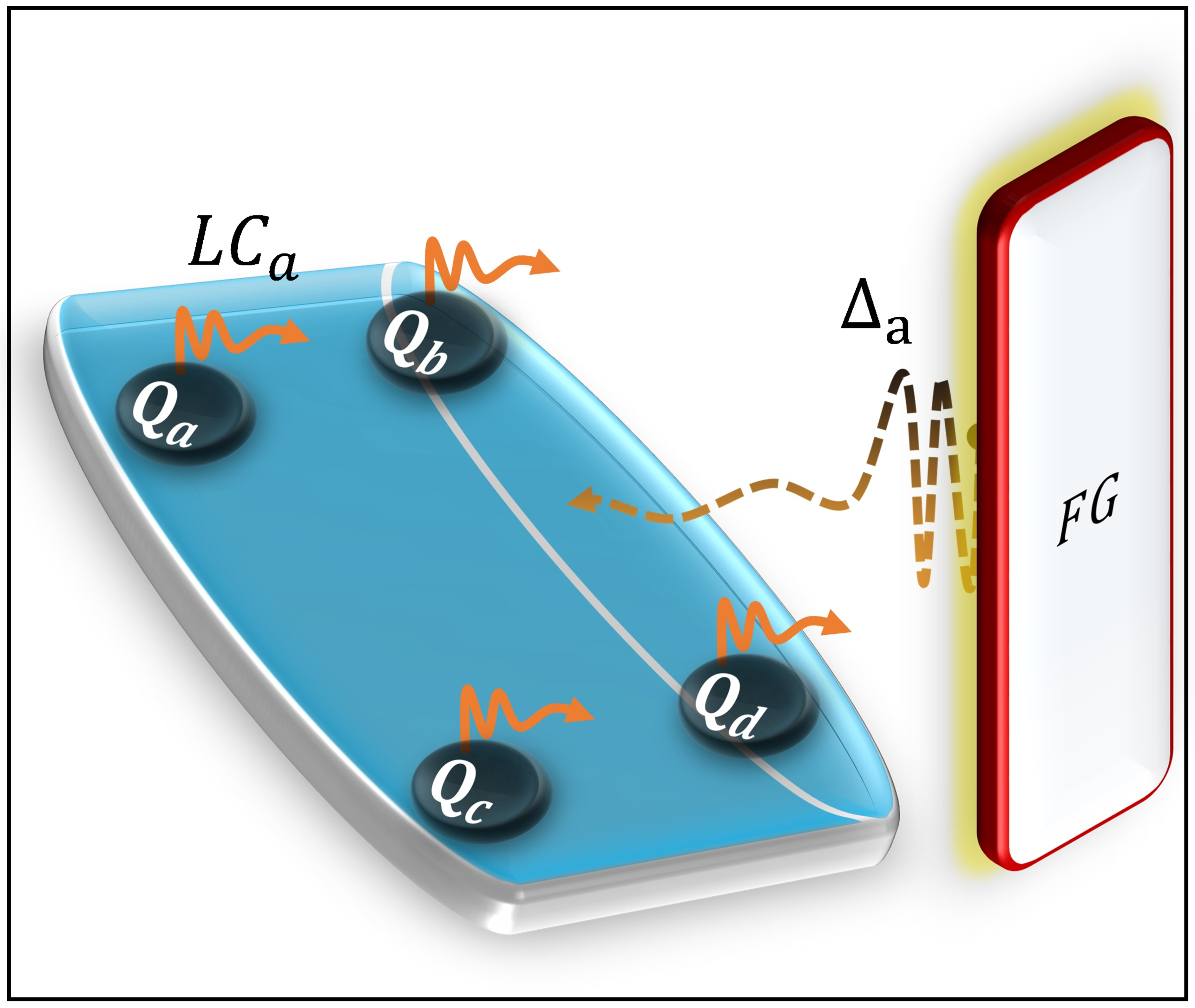} & \subfigimg[width=\linewidth]{(b)}{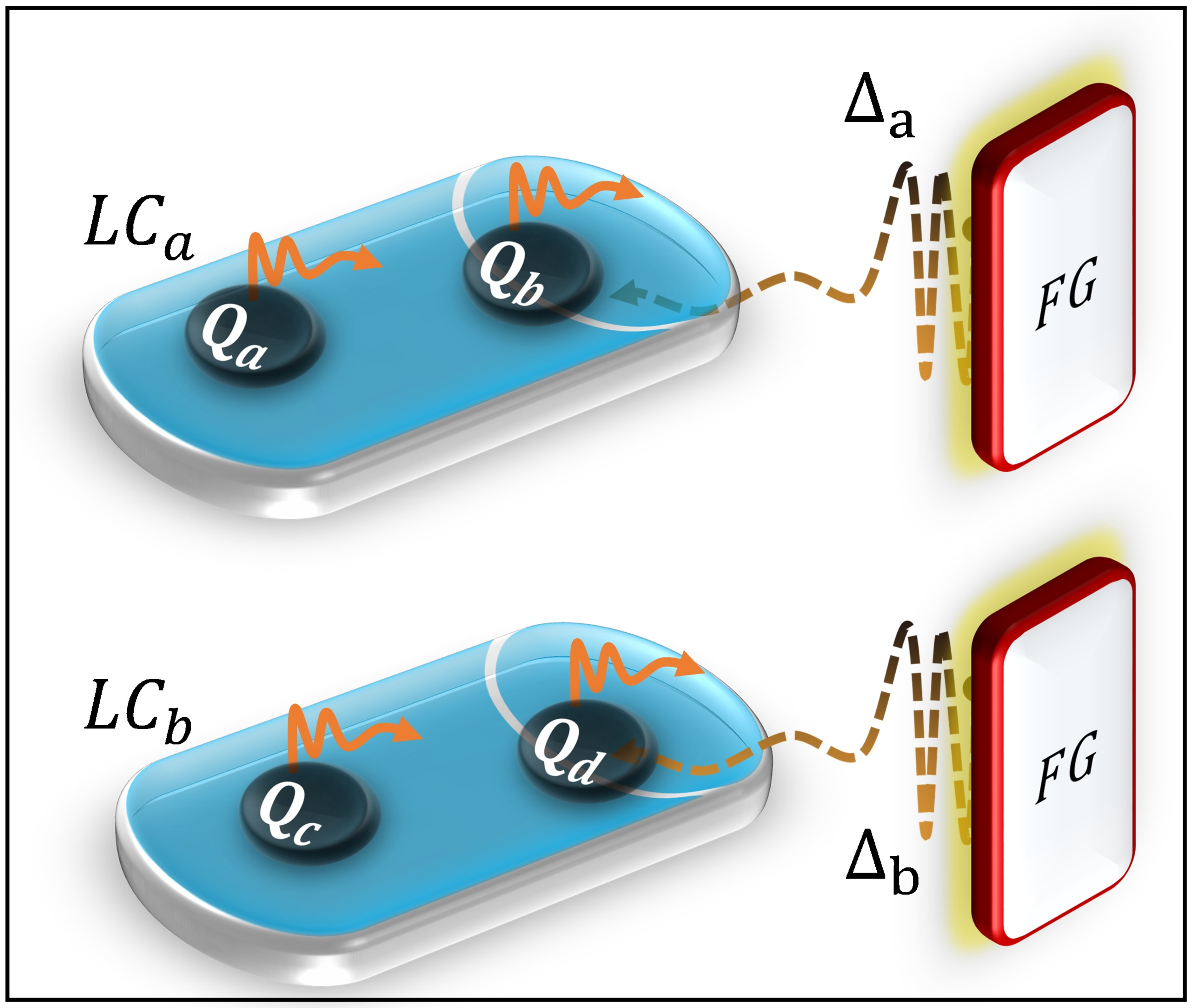} \\
						\subfigimg[width=\linewidth]{(c)}{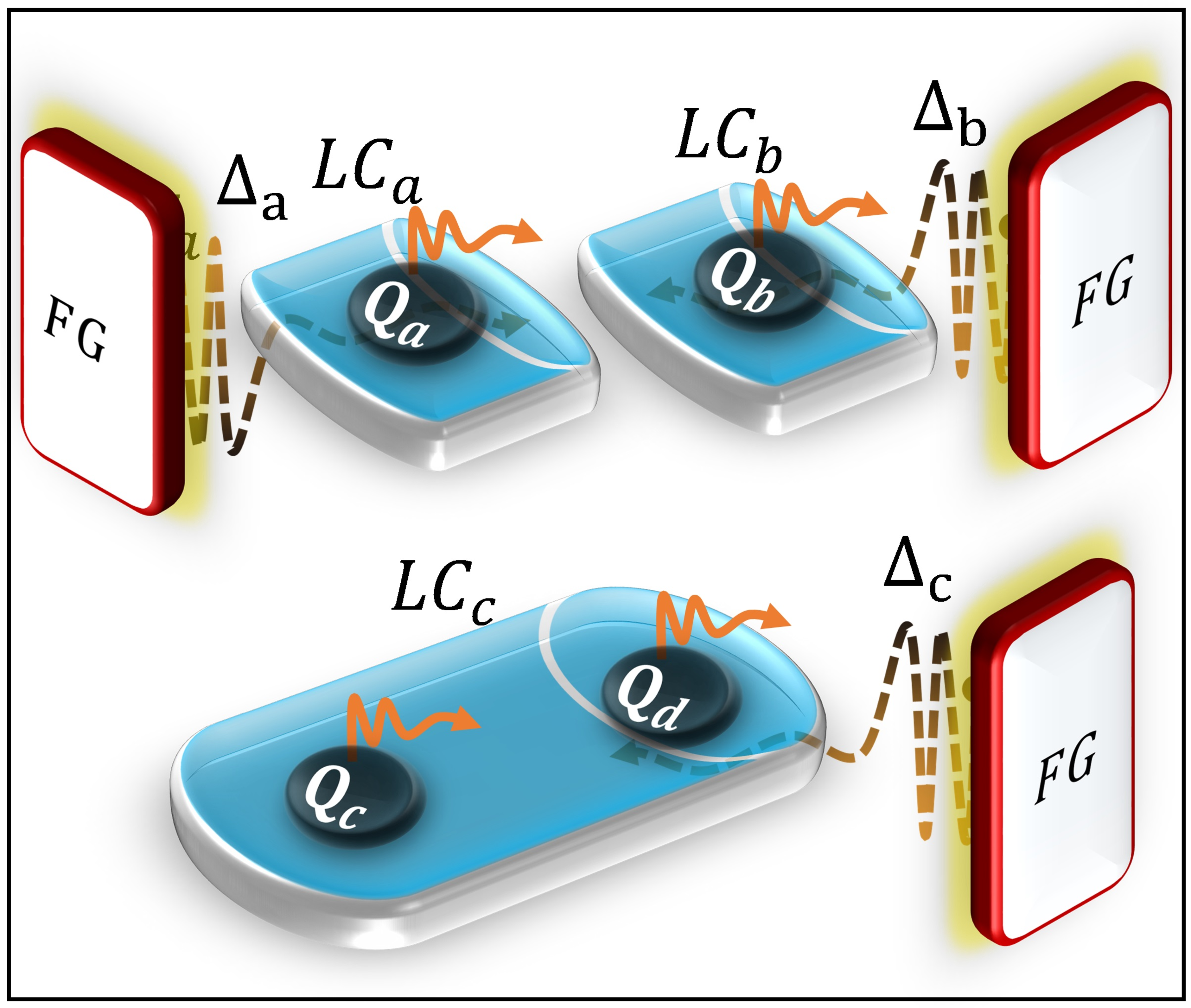} & \subfigimg[width=\linewidth]{(d)}{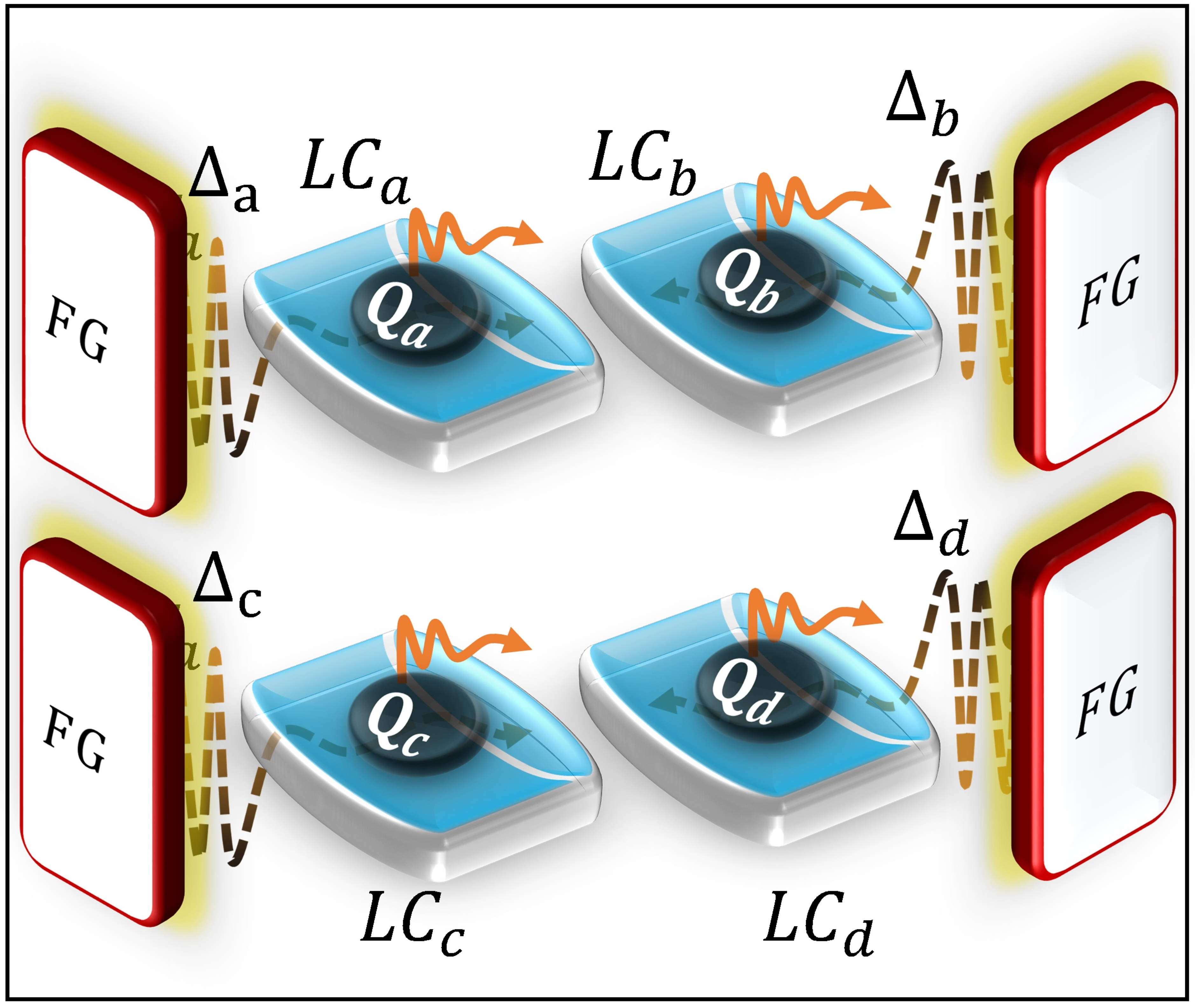} 
		\end{tabular}
\caption{Upper panel: Schematic diagram shows the present configurations used in this paper for four non-interacting maximally entangled qubits, $Q_a$, $Q_b$, $Q_c$ and $Q_d$ in local fields $LC_n$ where, $n \in \{a, b, c, d \}$ i.e., for common local channel-qubit (a) and bipartite local channel-qubit (b) configurations influenced by fractional Gaussian noise sources shown by the square like boxes named as FG. The black-brown wavy line shows the action of FG sources on the dynamics of qubits. $\Delta_n$ with $n \in \{a, b, c, d \}$ represents the stochastic parameter of the classical fields linking the noisy effects of FG sources with the phase factor of the system and local channel-qubit. The brownish wavy lines originating from qubits shows the related dynamics. Lower panel: Same as upper panel but for tripartite local channel-qubit (c) and independent local channel-qubit (d) configurations.}
\label{fig0}
	\end{figure}	
Our physical model comprises four identical non-interacting qubits with equivalent energy splitting $e_n$ coupled to classical fluctuating channels in four different configuration models. In the first case, all the four qubits are subjected to a single classical environment and are known as the CLCQ configuration. In the second and third situations, we present BSE and TLCQ configurations, where the system is coupled with two and three independent classical channels, respectively. In the final case, each qubit is exposed to an independent classical channel and is named as an ILCQ configuration (see Fig. \ref{fig0}). Note that each channel comprises a single FG noise source. The stochastic Hamiltonian governs the current physical model, which reads as \cite{53}:
\begin{equation}
H_{abcd}(t)=H_a(t) \otimes I_b \otimes I_c \otimes I_d+I_a \otimes H_b(t) \otimes I_c\otimes I_d+I_a \otimes I_b \otimes H_c(t)\otimes I_d+I_a \otimes I_b\otimes I_c \otimes H_d(t),\label{hmm}
\end{equation}
where $ H_n(t)$ is the Hamiltonian for a single qubit and is written as $ H_n(t)=e_n I+\omega \Delta_n(t)\sigma^x$ with $n \in \{a,b,c,d\}$. Here, $I$ and $\sigma^x$ are the identity and Pauli matrices operating on the qubit sub-spaces. $\Delta(t)$ is the classical stochastic parameter characterizing the noise, while $\omega$ is the coupling constant for the qubit-channel interaction. Next, the time evolved density matrix for the system is obtained by \cite{rr}:
\begin{equation}
\rho_{abcd}(t)=U_{abcd}(t)\rho_0 U_{abcd}(t)^{\dagger},\label{time evolved density matrix}
\end{equation}
where, $U_{abcd}(t)=\exp[-\imath\int^{t}_{t_0}H_{abcd}(s)ds]$ and is called time-unitary operation with $\hbar=1$ and $\rho_0$,  the initial density matrix for the system of four qubits and is given by \cite{dd}:
\begin{equation}
\rho_0 = \frac{(1-p)}{16}I_{(16\times 16)}+p\vert GHZ\rangle\langle GHZ\vert,\label{Maximal}
\end{equation}
with $0\leq p \leq 1$ and $ \vert GHZ\rangle $ is the four qubit maximally entangled Greenberger-Horne-Zeilinger state and is defined as $ \vert GHZ\rangle=\frac{1}{\sqrt{2}}(\vert 0000 \rangle + \vert 1111 \rangle)$.
\subsection{Fractional Gaussian noise}
We will cover the complete FG noise application in this section. To include the stochastic process, the $\beta$-function, which adds the noisy effects to the dynamics of the system, must be described, which read as \cite{56, 55}:
\begin{equation}
\beta(t)=\int_0^t \int_0^t dx dx^\prime K(x-x^\prime).\label{beta function}
\end{equation}
The auto-correlation function of the noise, that introduces the noise to the system phase in the local fields, is written as:
\begin{equation}
K_{F_G}(t-t^{\prime})=\frac{1}{2}(|t|^{2H}+|t^{\prime}|^{2H}-|t-t^{\prime}|^{2H}).\label{auto-correlation fn}%
\end{equation}
Now, one can get the corresponding $\beta$-function by substituting the auto-correlation function from Eq.\eqref{auto-correlation fn} into Eq. \eqref{beta function} as:
\begin{equation}
\beta_{FG}(\tau)=\frac{\tau^{2H+2}}{2H+2},\label{A13}%
\end{equation}
where, $ H $ is known as the Hurst parameter, which ranges between $0$ and $1$ \cite{57}. By averaging the elements of the time evolved density matrix over the noise determine the resulting dephasing effects and can be computed as $\langle \exp[\pm \imath k\phi_n(t)]  \rangle$=$\langle \exp[-\frac{1}{2}\beta_{FG}(\tau)(k)^2] \rangle$ with $\phi_n(t)=k \omega \Delta_n(t)$. When FG noise is introduced to CLCQ configuration, the corresponding final density matrix can be obtained by \cite{54}:
\begin{equation}
\rho_{CLCQ}(\tau)= \left\langle U_{abcd}(t)\rho_0  U_{abcd}(t)^{\dagger} \right\rangle_{\phi_a},\label{final density matrix of CSE}
\end{equation}
where, $\phi_a=\phi_b=\phi_c=\phi_d$. In the case of BLCQ configuration, the final density matrix can be obtained by:
\begin{equation}
\rho_{BLCQ}(\tau)= \left\langle \left\langle U_{abcd}(t)\rho_0  U_{abcd}(t)^{\dagger} \right\rangle_{\phi_a} \right\rangle_{\phi_b},\label{final density matrix of BSE}
\end{equation}
where we assume $\phi_c=\phi_a$ and $\phi_d=\phi_b$. Similarly, for the TLCQ configuration, the final density matrix can be computed by:
\begin{equation}
\rho_{TLCQ}(\tau)= \left\langle \left\langle \left\langle U_{abcd}(t)\rho_0  U_{abcd}(t)^{\dagger} \right\rangle_{\phi_a}\right \rangle_{\phi_b} \right\rangle_{\phi_c},\label{final density matrix of TSE}
\end{equation}
where we assume $\phi_d=\phi_c$. Finally, for the ILCQ configuration, the final density matrix is obtained by:
\begin{equation}
\rho_{ILCQ}(\tau)= \left\langle \left\langle \left\langle \left\langle U_{abcd}(t)\rho_0  U_{abcd}(t)^{\dagger} \right\rangle_{\phi_a} \right\rangle_{\phi_b} \right\rangle_{\phi_c} \right\rangle_{\phi_d}.\label{final density matrix for ISE}
\end{equation}
In the above equations, $\langle \Psi \rangle$ indicates the average over all the possible values of the FG noise phase applied on the state $\Psi$.
\section{Results and discussions}\label{Results and Discussions}
The major results for the dynamics of entanglement witness, negativity, purity, and mixedness for a system of four qubits initially prepared as mixed entangled GHZ state are given in this section for each qubit-noise configuration. However, in the current case, we only focus on maximally entangled four qubit GHZ state while setting $ p=1$ in Eq. \eqref{Maximal}.
\subsection{Common local channel-qubit configuration}
The dynamics of quantum correlation and coherence under $FG$ noise for the four qubit GHZ state are examined in this section. We focus on the coupling of all four qubits to a single classical fluctuating channel, whose final density matrix can be obtained by using Eq. \eqref{final density matrix of CSE}. The analytical relations for entanglement witness, N-partite negativity, purity, and von Neumann entropy measures are given in Eqs. \eqref{ewo},  \eqref{negativity}, \eqref{purity}, and \eqref{decoherence} can be put into the following forms:
\begin{align}
\text{ER}_{CLCQ}(\tau)&=\frac{1}{32} \left(e^{-32 \beta }+12 e^{-8 \beta }+3\right),&\\
\text{NY}_{CLCQ}(\tau)&=\frac{1}{32} \left( \sum^{7}_{i=1}\mathcal{A}_i\right) ,&\\
\text{PY}_{CLCQ}(\tau)&=\frac{1}{32} \left(19+e^{-64 \beta }+12 e^{-16 \beta }\right),&\\
\text{VE}_{CLCQ}(\tau)&=-\mathcal{A}_8-\frac{1}{8} (\mathcal{A}_9) \log\left[\frac{1}{16} \mathcal{A}_9\right].&
\end{align}
\begin{figure}[ht]
\includegraphics[width=0.49 \textwidth]{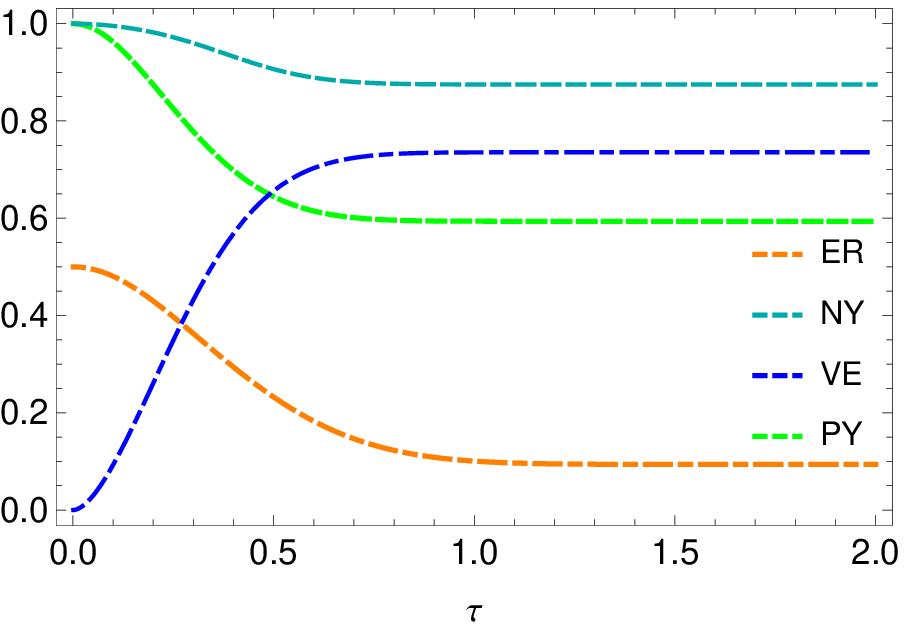}~~ \centering% ~~
\includegraphics[width=0.49 \textwidth]{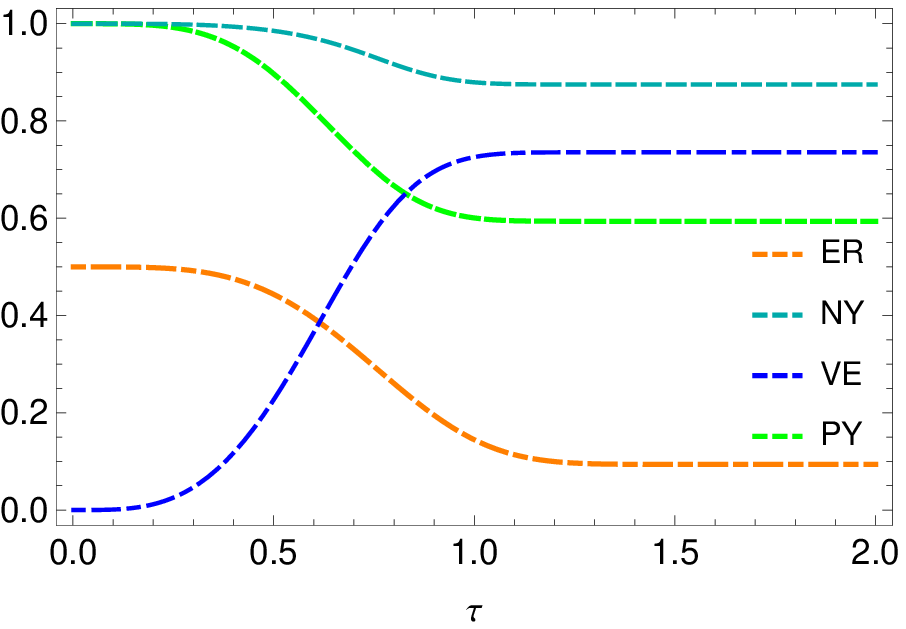} \put(-503, 0){(a)}
\put(-240,0){(b)}\\
\caption{Influence of fractional Gaussian noise on the dynamics of entanglement witness operation, negativity, purity and von Neumann entropy in multipartite GHZ state subjected to common local channel-qubit configuration with $H=0.01$ (a) and $H=0.9$ (b).}
\label{F1}
\end{figure}

Fig. \ref{F1} depicts the time evolution of entanglement, purity and mixedness in a four-qubit maximally entangled GHZ state coupled to a single source of FG noise. Under the influence of FG noise, the ER, NY, and PY remained decreasing functions of entanglement, coherence and purity. In contrast, the VE measure remained an increasing function of mixedness and decoherence. This demonstrates that FG noise degrades multipartite entanglement, coherence and purity while causing mixedness in the system. According to ER and NY, entanglement does not completely disappear, making the CLCQ configuration an excellent resource for extended memory properties while employing a four-qubit GHZ state. In comparison, the rate of disentanglement differs from the rate of decoherence and mixedness in the four-qubit state. As seen that the speed of mixedness in the four-qubit GHZ state is faster than that of the entanglement, coherence and purity decay. This can also be interpreted that because of the noise action of FG noise, the GHZ state becomes mixed, causing the state to be decoherent and lose entanglement. For a quantum system, entanglement can be protected by designing the system-channel coupling with a single noise source, contributing to improving the efficiency of quantum mechanical protocols such as witnessed in the current case. The saturation of entanglement, coherence and purity in four-qubit state at different time intervals shows that the current FG noise has a different impact on these phenomena. All the metrics, ER, NY, VE and PY, have shown a monotonic decline of the correlations between the qubits rather than any rebirths. This confirms that in the CLCQ configuration under FG noise, the system loses information that is not recoverable. As shown in \cite{37,38,48}, the overall dephasing effects of the FG noise differ completely from non-Gaussian noises, such as random telegraph and static noise. Under the FG noise, the previously studied bipartite Bells state has achieved full separability \cite{56}, however, entanglement is not lost for the current four-qubit state. Thus, ensuring the less fragile nature of the entanglement, coherence and purity in the current four-qubits GHZ class state against the FG noise. At the upper bound of $H$, entanglement and purity remained more robust. Thus, increasing the $H$ exponent improves entanglement, coherence and purity survival time; nonetheless, the total FG noisy dephasing effects are unavoidable. The metrics involved showed strong connections. For example, the rise in VE functions speed can be traced back to the declining speed of the ER, NY, and PY functions. 
\subsection{Bipartite local channel-qubit configuration}\label{GHZ state when coupled to BLCQ configuration}
This section examines the dynamics of entanglement, purity and mixedness functions when the four qubits are coupled with two separate channels, each with an equal number of qubits. The final density matrix of the current configuration is given by Eq. \eqref{final density matrix of BSE}. The analytical results of entanglement, purity and mixedness are obtained using Eqs. \eqref{ewo}, \eqref{negativity}, \eqref{purity}, and \eqref{decoherence} and take the forms:

\begin{align}
\text{ER}_{BLCQ}(\tau)&=\frac{1}{16} \left(e^{-16 \beta }+e^{-8 \beta }+8 e^{-4 \beta }+e^{-2 \beta }-3\right),&\\
\text{NY}_{BLCQ}(\tau)&=\frac{1}{192} e^{-16 \beta }\left( \sum^{12}_{i=1}\mathcal{B}_i\right) ,&\\
\text{PY}_{BLCQ}(\tau)&=\frac{1}{16} e^{-32 \beta } \left(1+e^{16 \beta }+8 e^{24 \beta }+e^{28 \beta }+5 e^{32 \beta }\right),&\\
\text{VE}_{BLCQ}(\tau)&=-\mathcal{B}_{13}-\mathcal{B}_{14}-\frac{1}{16} (\mathcal{B}_{15}) \log\left[\frac{1}{16}\mathcal{B}_{15}\right].&
\end{align}

\begin{figure}[ht]
\includegraphics[width=0.49 \textwidth]{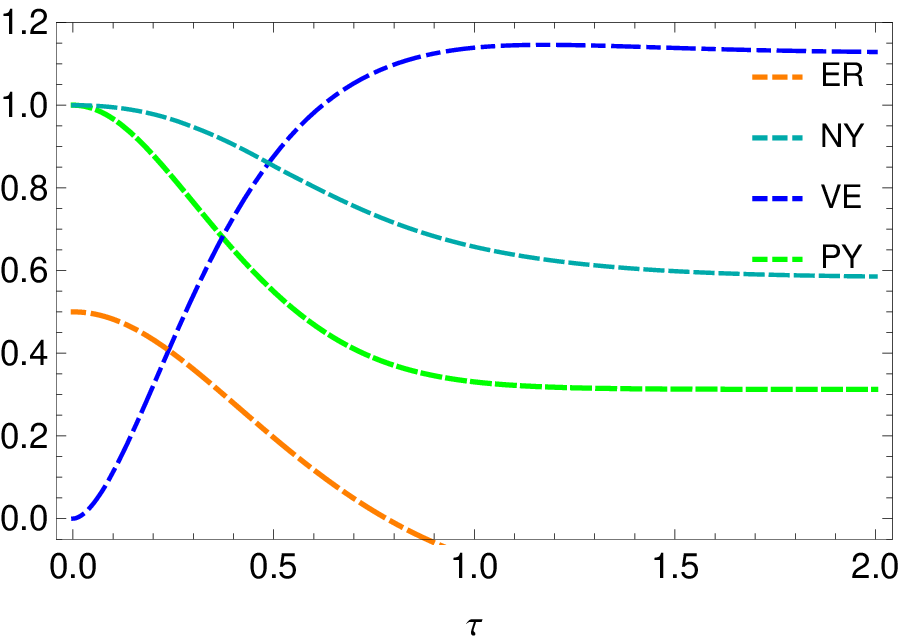}~~ \centering% ~~
\includegraphics[width=0.49 \textwidth]{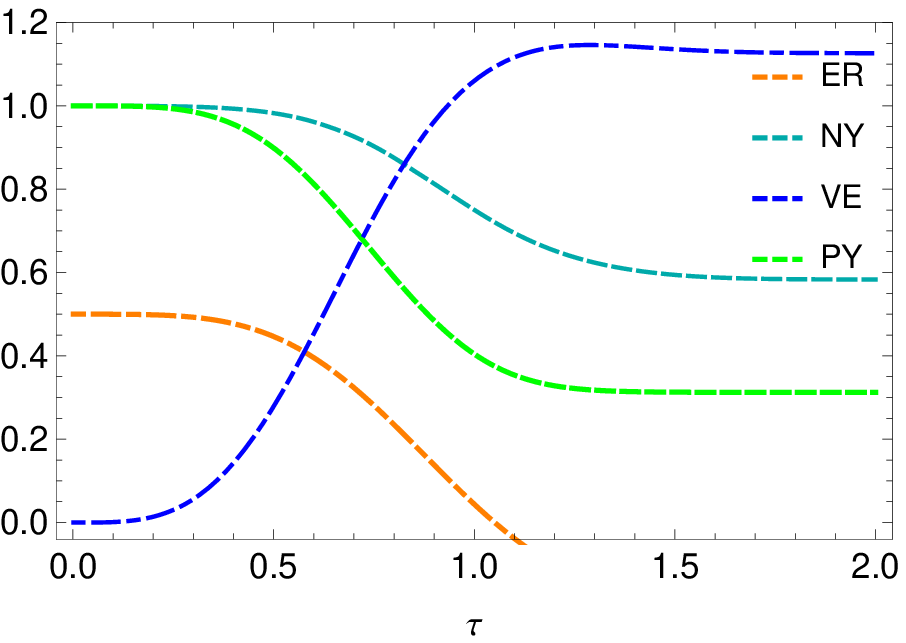}\put(-503, 0){(a)}
\put(-240,0){(b)}\\
\caption{Influence of fractional Gaussian noise on the dynamics of entanglement witness operation, negativity, purity and von Neumann entropy in multipartite GHZ state subjected to bipartite local channel-qubit configuration with $H=0.01$ (a) and $H=0.9$ (b).}
\label{F2}
\end{figure}
In Fig. \ref{F2}, we plot the time evolution of ER, NE, PY, and VE to examine the characteristic behaviour of the entanglement, purity and mixedness in four-qubit maximally entangled GHZ class state exposed to FG noisy bipartite local channels. When the BLCQ configuration is employed, the entanglement is more brittle to the FG noise. The state faces greater entanglement decay besides large mixedness in the system of four entangled qubits, according to ER, NY, PY, and VE measures. The curves of the current metrics approach their final asymptotic values in a full exponential manner, showing no rebirths. Non-Gaussian noises such as random telegraph and static noise, in contrast to FG noise, have been discovered to support strong entanglement sudden death and birth revivals \cite{37,38,48}. This implies a strong propensity of the FG noise to limit information exchange between the system and associated bipartite local channels. The qualitative dynamics was consistent across ER, NE, PY, and VE measures, and convey that for the rise in mixedness in the system, the entanglement, coherence and purity consequently get reduced. Although, the mixedness rate in the four entangled qubits leads to the relevant disentanglement and purity decay rate. In addition, the ER measure is more susceptible to the external noisy effects and shows a gradual death, on the other hand, NY remained a non-zero function. This comprehends the strong measurement capacity of NY measure to capture entanglement in the four qubits, which was not detected by ER measure. It is noticeable that, in the current BLCQ configuration, the four-qubit entangled state gets partially disentangled rather than becoming an absolute product state. This proves the resourcefulness of the four-qubit GHZ like state to remain largely tolerant of the noisy actions of the local channels. On the other hand, the results in \cite{56} for a system of two qubits employing quantum negativity and quantum discord show complete separability. Compared to the CLCQ configuration, the entanglement and coherence loss rate seems faster and greater in the current case. The reason is that the system is distributed in more than one local channel, causing faster mixedness in the maximally entangled four-qubit state.
\subsection{Tripartite local channel-qubit configuration}
In the current configuration, the four qubits are linked to three separate channels described by FG noise. The final density matrix is given by Eq. \eqref{final density matrix of TSE}. To provide details on entanglement and purity dynamics, we use entanglement witness, quantum negativity, purity, and von Neumann entropy measures from Eqs. \eqref{ewo}, \eqref{negativity}, \eqref{purity}, and \eqref{decoherence}. The corresponding analytical results and numerical simulations are as under:
\begin{align}
\text{ER}_{TLCQ}=&\frac{1}{16} e^{-12 \beta } \left(11 e^{8 \beta }+e^{10 \beta }-5 e^{12 \beta }+1\right),\\
\text{NY}_{TLCQ}=&\frac{1}{192} \left(-192+\sum^{10}_{i=1}\mathcal{C}_i\right),\\
\text{PY}_{TLCQ}=&\frac{1}{16} \left(5+8 e^{-8 \beta }+3 e^{-4 \beta }\right),\\
\text{VE}_{TLCQ}=&-\mathcal{C}_{11}-\mathcal{C}_{12}-\mathcal{C}_{12}-\frac{1}{8}\mathcal{C}_{14}\log\left[\frac{1}{16} \mathcal{C}_{14} \right].
\end{align}

\begin{figure}[ht]
\includegraphics[width=0.49 \textwidth]{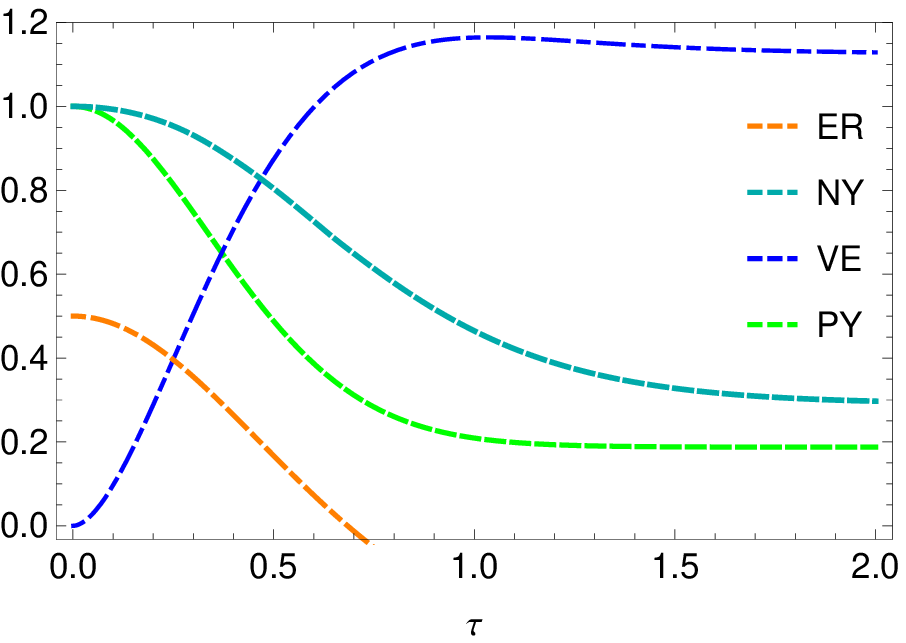}~~ \centering% ~~
\includegraphics[width=0.49 \textwidth]{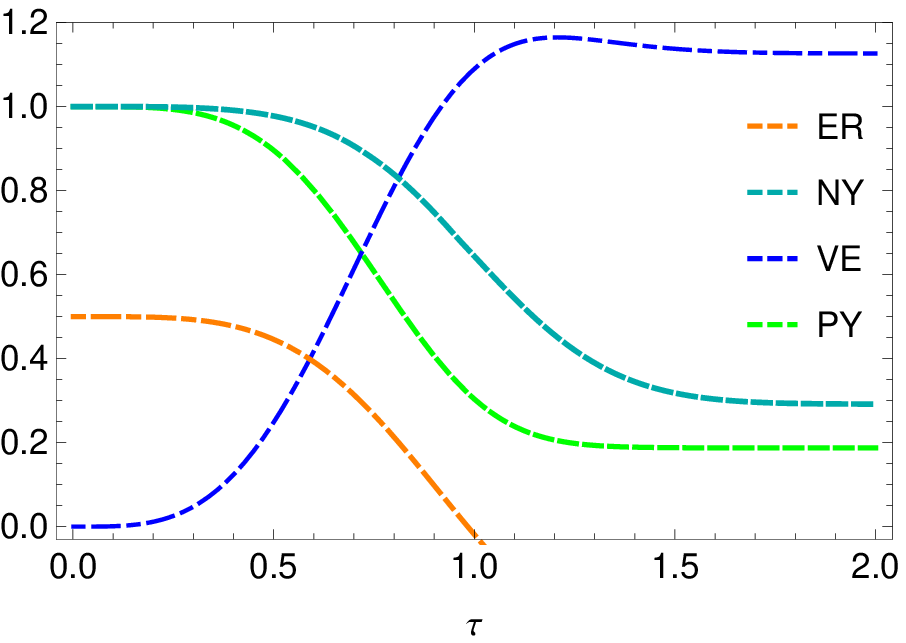} \put(-503, 0){(a)}
\put(-240,0){(b)}\\
\caption{Influence of fractional Gaussian noise on the dynamics of entanglement witness operation, negativity, purity and von Neumann entropy in multipartite GHZ state subjected to tripartite local channel-qubit configuration with $H=0.01$ (a) and $H=0.9$ (b).}\label{F3}
\end{figure}
In Fig. \ref{F3}, the key findings obtained for the dynamics of the entanglement, and mixedness in multipartite GHZ state when coupled to TLCQ configuration influenced by FG noise are reported. We noticed that the ER, NY, and PY functions in the case of a four-qubit GHZ like state remained decreasing functions of entanglement and purity in time under the influence of the FG noisy effects in the three local channels. VE function, on the other hand, has remained an increasing function of mixedness in four qubits. The entanglement, coherence and purity decline was monotonic, and no revivals of quantum phenomena were observed. This implies that the current Gaussian noisy local channels do not allow for the entanglement sudden death and birth phenomena. Thus, the information flow to the system and the conversion of the free states into resource states is not possible in the current local channels. The fact that the current results differ from those obtained for the BLCQ and CLCQ configurations is significant. In the current case, entanglement, coherence and purity of the four-qubit state were more fragile and suffered a greater loss. Although the general qualitative dynamics appear to be similar, the quantitative features measured for the current metrics are different, such as entanglement, purity and coherence preservation time. Besides, ER measure has shown a gradual death, however, the NY measure remained a non-zero function of entanglement. This confirms that in the case of four qubits, the NY measure is one of the resourceful entanglement monotones. The growing speed of the VE measure seems higher than compared to other involved measures, suggesting faster mixedness in the system, which results lately in disentanglement and decoherence. With higher $H$ values, there is more robustness in the entanglement and coherence, which improves the system's memory attributes. Instead of being becoming fully separable, the state becomes disentangled, mixed and decohere partially in the TLCQ configuration because of the dephasing effects caused by the FG noise phase. When such entangled states are combined with non-Gaussian noises in classical environments, the results are drastically different \cite{37, 38, 48, 53,rr}.  
\subsection{Independent local channel-qubit configuration}
When each of the four qubits is connected to an independent classical channel, the dynamics of entanglement and purity under FG noise are described in this section. Eq. \eqref{final density matrix for ISE} represents the current configuration's final density matrix. We use entanglement witness, quantum negativity, purity, and von Neumann entropy measures given in Eqs. \eqref{ewo}, \eqref{negativity}, \eqref{purity}, and \eqref{decoherence} to provide details on the entanglement, purity and mixedness dynamics, which are:
\begin{align}
\text{ER}_{ISE}(\tau)=&\frac{1}{8} \left(e^{-8 \beta }+6 e^{-4 \beta }-3\right),\\
\text{NY}_{ISE}(t)=&\frac{1}{8} e^{-8 \beta} (\mathcal{D}_1+\mathcal{D}_2),\\
\text{PY}_{ISE}(t)=&\frac{1}{8} \left(1+e^{-16 \beta }+6 e^{-8 \beta }\right),\\
\text{VE}_{ISE}(t)=&-\mathcal{D}_3-\mathcal{D}_4-\mathcal{D}_4.
\end{align}
\begin{figure}[ht]
\includegraphics[width=0.47 \textwidth]{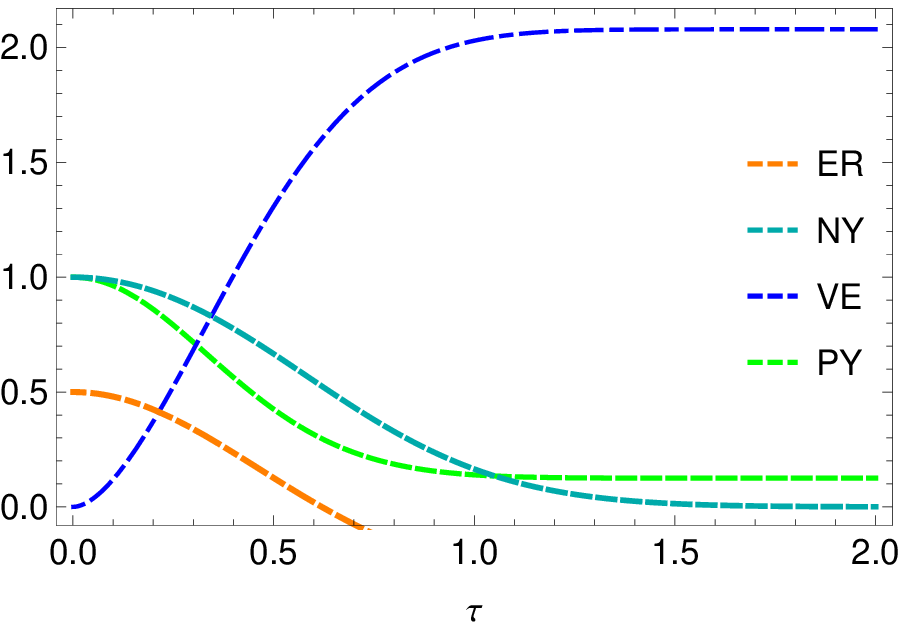}\centering%
\includegraphics[width=0.47 \textwidth]{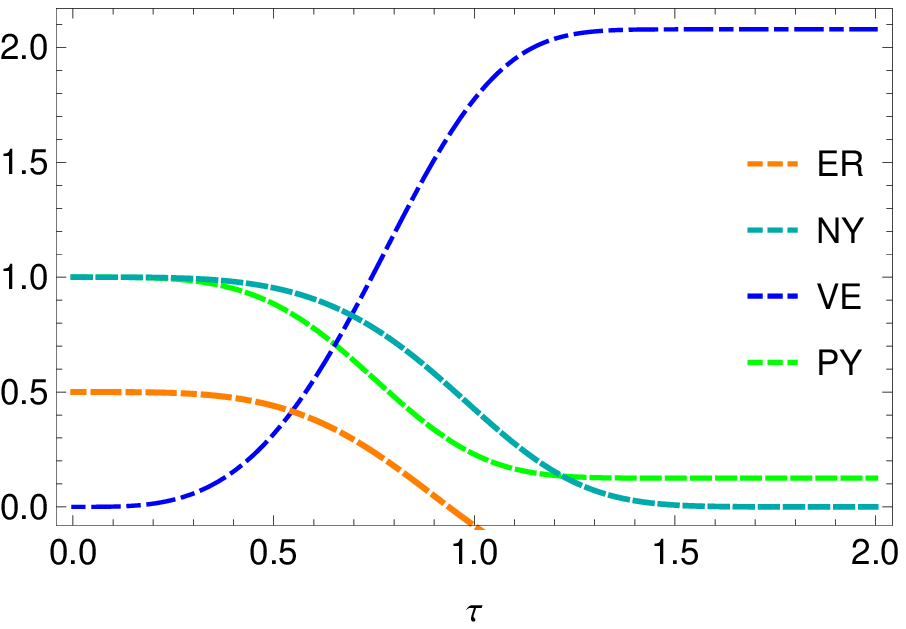} \put(-503, 0){(a)}
\put(-240,0){(b)}\\
\caption{Influence of fractional Gaussian noise on the dynamics of entanglement witness operation, negativity, purity and von Neumann entropy in multipartite GHZ state subjected to independent local channel-qubit configuration with $H=0.01$ (a) and $H=0.9$ (b).}\label{dynamics of GHZ state under ISE}
\end{figure}
Fig. \ref{dynamics of GHZ state under ISE} investigate the dynamics of entanglement, purity and mixedness in four qubit GHZ class state when coupled to ILCQ configuration driven by FG noise. In line with previous findings, ER, NY and PY remained declining functions of entanglement and purity in time. On the other hand, VE remained an increasing function of mixedness in the four qubits. The detrimental degrading effects are completely different quantitatively, with the resulting loss being significantly higher when four qubits are distributed among four independent local channels. The entangled sudden death and birth phenomenon is completely halted, and the decay is completely monotone. Because there is no flow of lost information from the channel back to the system, this translates to irreversible decay. We found that the independent noisy channels and the related decay have an intrinsic relationship, where both are directly related. Aside from that, the slopes show a reduced loss as $H$ increases, signifying that the system's memory characteristics have improved. This contradicts the findings of previous studies for a variety of noisy parameters, as given in \cite{37,38,48,53,rr,55,ff,ii,jj} where the increasing noise parameters caused the quantum correlations or coherence decay to speed up. The complete disentanglement and higher mixedness rates remained different at a comparable time for the four qubits. Because of the mixedness generated by the coupling of the system with the environment, the system becomes unentangled as shown by ER, NY, PY and VE. This arrangement could be useful for quantum mechanical protocols that require a great deal of channel capacity. All the metrics produced the same results, suggesting that the results are consistent and valid and that they have a strong relationship.

\par
In this study, we looked at quantum correlations, purity, and mixedness using the ER, NY, PY, and VE functions. A thorough qualitative and quantitative investigation was carried out on four qubits exposed to local noisy channels. The ER and NY measures agree only when the four qubits are coupled to a single local channel, ensuring partial loss of entanglement and purity. In all other cases, when compared to the NY measure, the ER measure remained extremely fragile, quickly reaching zero and demonstrating complete disentanglement of the four-qubit system. However, the NY measure remained a non-zero entanglement function in all cases of the different channels except in the ILCQ configuration. Thus, NY remained a more appropriate entanglement monotone in the case of four-qubit entanglement. Besides, the PY measure agrees with ER measure and shows that the system only remained pure in the CLCQ model of the channels, and becomes mixed in all other cases. However, in the rest of the channels types, we see that the loss in purity is increasing, meaning that the state is still pure. This statement can be verified by looking into the curves obtained for the VE measure, which constantly arise for the increasing number of coupled channels, ensuring that only in ILCQ channels the mixedness is enough high, which indeed will cause the state to become completely mixed.
\par

\section{Conclusion}\label{Conclusion}
Quantum correlations and mixedness must be thoroughly investigated before they can be put into practice for quantum information processing to be successful. When using quantum systems for non-local correlations and purity, quantum mechanical protocols become inefficient due to external decoherence effects. Quantum state spaces must be optimized for noise parameters in order to avoid these negative consequences. This requires precise characterization of various noises and optimal noise parameter settings. Entanglement and purity are studied for a maximally entangled GHZ like state. For quantum information processing purposes, we employ local noisy channels for information processing through four qubits. We assume the local channels are driven by discrete Brownian motion, causing FG noise. The combination of four qubits, channels and noise sources is studied in four different models, namely, common local channel-qubit, bipartite local channel-qubit, tripartite local channel-qubit and independent local channel-qubit configurations. 

\par

Entanglement and purity decay monotonically under the influence of FG noise. This means that information exchange and the conversion of free into resources states is an unavoidable defect in the FG noise local channels. In the current local channels, the four-qubit state successfully transmitted correlations between the qubits either for an indefinite or definite time. In the common, bipartite and tripartite local channel-qubit models, the four-qubit system remained partially entangled, while in the independent one, the system becomes completely separable. The amount of decay has been noticed to be regulated by the number of channels and both directly increase with each other. Thus, to implement effective quantum information processing protocols, it is useful to design configurations with minimal channels.
\par

The Hurst exponent $(H)$ has been observed to have the unusual property of initially protecting entanglement and purity. This directly opposes the properties of nearly all the noisy parameters described in \cite{37,38,48,53,55, rr, gg, jj, dd, ff, ii, ATTA-GN,  ATTA-PLFG}. Besides, we found no exact solution for avoiding the relative noisy effects in any potential situation. Hence, quantum correlations, purity and information initially encoded in any kind of quantum system subjected to FG noise will ultimately face decaying effects. However, lowering the temperature of the local channels can contribute to lowering the random motion of the particles and hence, reducing the FG noise effects.
\par

The current four qubit GHZ-class states were described as being a good resource for quantum information processing that can withstand noisy effects for both finite and infinite intervals, depending on the number of local channels involved. The four-qubit state has also been observed to be more entangled, coherent and pure, compared to the bipartite and tripartite states. Because it has a larger capacity for storing quantum information, the four-qubit state is likely to be more efficient than other quantum systems. As a result, the proper selection of quantum systems, system-channel couplings, the number of channels, and noise parameter tuning can improve the performance of quantum information processing protocols.\\
\section{Appendix}
In this section, we give the details of the final density matrices for the four qubit GHZ state when coupled to different classical environmental configurations driven by FG noise.
By using the Eq. \eqref{final density matrix of CSE}, the corresponding final density matrix for the CLCQ configuration takes the form as:
\begin{align}
\rho_{CSE}(\tau)=\left[
\begin{array}{cccccccccccccccc}
 \mathcal{X}_1 & 0 & 0 & \mathcal{X}_2 & 0 & \mathcal{X}_2 & \mathcal{X}_2 & 0 & 0 & \mathcal{X}_2 & \mathcal{X}_2 & 0 & \mathcal{X}_2 & 0 & 0 & \mathcal{X}_1 \\
 0 & \mathcal{X}_3 & \mathcal{X}_3 & 0 & \mathcal{X}_3 & 0 & 0 & \mathcal{X}_3 & \mathcal{X}_3 & 0 & 0 & \mathcal{X}_3 & 0 & \mathcal{X}_3 & \mathcal{X}_3 & 0 \\
 0 & \mathcal{X}_3 & \mathcal{X}_3 & 0 & \mathcal{X}_3 & 0 & 0 & \mathcal{X}_3 & \mathcal{X}_3 & 0 & 0 & \mathcal{X}_3 & 0 & \mathcal{X}_3 & \mathcal{X}_3 & 0 \\
 \mathcal{X}_2 & 0 & 0 & \mathcal{X}_4 & 0 & \mathcal{X}_4 & \mathcal{X}_4 & 0 & 0 & \mathcal{X}_4 & \mathcal{X}_4 & 0 & \mathcal{X}_4 & 0 & 0 & \mathcal{X}_2 \\
 0 & \mathcal{X}_3 & \mathcal{X}_3 & 0 & \mathcal{X}_3 & 0 & 0 & \mathcal{X}_3 & \mathcal{X}_3 & 0 & 0 & \mathcal{X}_3 & 0 & \mathcal{X}_3 & \mathcal{X}_3 & 0 \\
 \mathcal{X}_2 & 0 & 0 & \mathcal{X}_4 & 0 & \mathcal{X}_4 & \mathcal{X}_4 & 0 & 0 & \mathcal{X}_4 & \mathcal{X}_4 & 0 & \mathcal{X}_4 & 0 & 0 & \mathcal{X}_2 \\
 \mathcal{X}_2 & 0 & 0 & \mathcal{X}_4 & 0 & \mathcal{X}_4 & \mathcal{X}_4 & 0 & 0 & \mathcal{X}_4 & \mathcal{X}_4 & 0 & \mathcal{X}_4 & 0 & 0 & \mathcal{X}_2 \\
 0 & \mathcal{X}_3 & \mathcal{X}_3 & 0 & \mathcal{X}_3 & 0 & 0 & \mathcal{X}_3 & \mathcal{X}_3 & 0 & 0 & \mathcal{X}_3 & 0 & \mathcal{X}_3 & \mathcal{X}_3 & 0 \\
 0 & \mathcal{X}_3 & \mathcal{X}_3 & 0 & \mathcal{X}_3 & 0 & 0 & \mathcal{X}_3 & \mathcal{X}_3 & 0 & 0 & \mathcal{X}_3 & 0 & \mathcal{X}_3 & \mathcal{X}_3 & 0 \\
 \mathcal{X}_2 & 0 & 0 & \mathcal{X}_4 & 0 & \mathcal{X}_4 & \mathcal{X}_4 & 0 & 0 & \mathcal{X}_4 & \mathcal{X}_4 & 0 & \mathcal{X}_4 & 0 & 0 & \mathcal{X}_2 \\
 \mathcal{X}_2 & 0 & 0 & \mathcal{X}_4 & 0 & \mathcal{X}_4 & \mathcal{X}_4 & 0 & 0 & \mathcal{X}_4 & \mathcal{X}_4 & 0 & \mathcal{X}_4 & 0 & 0 & \mathcal{X}_2 \\
 0 & \mathcal{X}_3 & \mathcal{X}_3 & 0 & \mathcal{X}_3 & 0 & 0 & \mathcal{X}_3 & \mathcal{X}_3 & 0 & 0 & \mathcal{X}_3 & 0 & \mathcal{X}_3 & \mathcal{X}_3 & 0 \\
 \mathcal{X}_2 & 0 & 0 & \mathcal{X}_4 & 0 & \mathcal{X}_4 & \mathcal{X}_4 & 0 & 0 & \mathcal{X}_4 & \mathcal{X}_4 & 0 & \mathcal{X}_4 & 0 & 0 & \mathcal{X}_2 \\
 0 & \mathcal{X}_3 & \mathcal{X}_3 & 0 & \mathcal{X}_3 & 0 & 0 & \mathcal{X}_3 & \mathcal{X}_3 & 0 & 0 & \mathcal{X}_3 & 0 & \mathcal{X}_3 & \mathcal{X}_3 & 0 \\
 0 & \mathcal{X}_3 & \mathcal{X}_3 & 0 & \mathcal{X}_3 & 0 & 0 & \mathcal{X}_3 & \mathcal{X}_3 & 0 & 0 & \mathcal{X}_3 & 0 & \mathcal{X}_3 & \mathcal{X}_3 & 0 \\
 \mathcal{X}_1 & 0 & 0 & \mathcal{X}_2 & 0 & \mathcal{X}_2 & \mathcal{X}_2 & 0 & 0 & \mathcal{X}_2 & \mathcal{X}_2 & 0 & \mathcal{X}_2 & 0 & 0 & \mathcal{X}_1
\end{array}
\right],
\end{align}
where
\begin{align*}
\mathcal{X}_1=&\frac{19}{64}+\eta_1+\eta_2, &  \mathcal{X}_2=&-\frac{1}{64}(1+\eta_3)+\eta_4,\\
\mathcal{X}_3=&\frac{1}{64}(1-\eta_5), & \mathcal{X}_4=&\frac{1}{64}(1+\eta_6)-\eta_7.\\
\end{align*}
By utilizing the Eq. \eqref{final density matrix of BSE}, final density matrix of the BLCQ configuration is given as:
\begin{align}
\rho_{BSE}(\tau)=\left[
\begin{array}{cccccccccccccccc}
 \mathcal{P}_1 & 0 & 0 & \mathcal{P}_2 & 0 & 0 & 0 & 0 & 0 & 0 & 0 & 0 & \mathcal{P}_2 & 0 & 0 & \mathcal{P}_1 \\
 0 & \mathcal{P}_3 & \mathcal{P}_3 & 0 & 0 & 0 & 0 & 0 & 0 & 0 & 0 & 0 & 0 & \mathcal{P}_3 & \mathcal{P}_3 & 0 \\
 0 & \mathcal{P}_3 & \mathcal{P}_3 & 0 & 0 & 0 & 0 & 0 & 0 & 0 & 0 & 0 & 0 & \mathcal{P}_3 & \mathcal{P}_3 & 0 \\
 \mathcal{P}_2 & 0 & 0 & \mathcal{P}_4 & 0 & 0 & 0 & 0 & 0 & 0 & 0 & 0 & \mathcal{P}_4 & 0 & 0 & \mathcal{P}_2 \\
 0 & 0 & 0 & 0 & \mathcal{P}_5 & 0 & 0 & \mathcal{P}_5 & \mathcal{P}_5 & 0 & 0 & \mathcal{P}_5 & 0 & 0 & 0 & 0 \\
 0 & 0 & 0 & 0 & 0 & \mathcal{P}_6 & \mathcal{P}_6 & 0 & 0 & \mathcal{P}_6 & \mathcal{P}_6 & 0 & 0 & 0 & 0 & 0 \\
 0 & 0 & 0 & 0 & 0 & \mathcal{P}_6 & \mathcal{P}_6 & 0 & 0 & \mathcal{P}_6 & \mathcal{P}_6 & 0 & 0 & 0 & 0 & 0 \\
 0 & 0 & 0 & 0 & \mathcal{P}_5 & 0 & 0 & \mathcal{P}_5 & \mathcal{P}_5 & 0 & 0 & \mathcal{P}_5 & 0 & 0 & 0 & 0 \\
 0 & 0 & 0 & 0 & \mathcal{P}_5 & 0 & 0 & \mathcal{P}_5 & \mathcal{P}_5 & 0 & 0 & \mathcal{P}_5 & 0 & 0 & 0 & 0 \\
 0 & 0 & 0 & 0 & 0 & \mathcal{P}_6 & \mathcal{P}_6 & 0 & 0 & \mathcal{P}_6 & \mathcal{P}_6 & 0 & 0 & 0 & 0 & 0 \\
 0 & 0 & 0 & 0 & 0 & \mathcal{P}_6 & \mathcal{P}_6 & 0 & 0 & \mathcal{P}_6 & \mathcal{P}_6 & 0 & 0 & 0 & 0 & 0 \\
 0 & 0 & 0 & 0 & \mathcal{P}_5 & 0 & 0 & \mathcal{P}_5 & \mathcal{P}_5 & 0 & 0 & \mathcal{P}_5 & 0 & 0 & 0 & 0 \\
 \mathcal{P}_2 & 0 & 0 & \mathcal{P}_4 & 0 & 0 & 0 & 0 & 0 & 0 & 0 & 0 & \mathcal{P}_4 & 0 & 0 & \mathcal{P}_2 \\
 0 & \mathcal{P}_3 & \mathcal{P}_3 & 0 & 0 & 0 & 0 & 0 & 0 & 0 & 0 & 0 & 0 & \mathcal{P}_3 & \mathcal{P}_3 & 0 \\
 0 & \mathcal{P}_3 & \mathcal{P}_3 & 0 & 0 & 0 & 0 & 0 & 0 & 0 & 0 & 0 & 0 & \mathcal{P}_3 & \mathcal{P}_3 & 0 \\
 \mathcal{P}_1 & 0 & 0 & \mathcal{P}_2 & 0 & 0 & 0 & 0 & 0 & 0 & 0 & 0 & \mathcal{P}_2 & 0 & 0 & \mathcal{P}_1
\end{array}
\right]
\end{align}
where
\begin{align*}
\mathcal{P}_1=&\frac{1}{32} \kappa_1 \left(1+\kappa_2)\right), & \mathcal{P}_2=&\frac{1}{32} \kappa_3 \left(\kappa_4\right),\\
\mathcal{P}_3=&\frac{1}{32}(-1+ \kappa_5\left(\kappa_6\right)),& \mathcal{P}_4=&\frac{1}{32} \kappa_7 \left(1+\kappa_8\right),\\
\mathcal{P}_5=&\frac{1}{32} \kappa_9 \left(-1+\kappa_{10} \text{Sinh}[\beta ]\right).
\end{align*}
For the TLCQ configuration, we can get the final density matrix by using the Eq. \eqref{final density matrix of TSE} as:
\begin{align}
\rho_{TSE}(\tau)=\left[
\begin{array}{cccccccccccccccc}
 \mathcal{Q}_1 & 0 & 0 & \mathcal{Q}_2 & 0 & \mathcal{Q}_2 & \mathcal{Q}_2 & 0 & 0 & \mathcal{Q}_2 & \mathcal{Q}_2 & 0 & \mathcal{Q}_2 & 0 & 0 & \mathcal{Q}_1 \\
 0 & \mathcal{Q}_3 & \mathcal{Q}_2 & 0 & \mathcal{Q}_2 & 0 & 0 & \mathcal{Q}_2 & \mathcal{Q}_2 & 0 & 0 & \mathcal{Q}_2 & 0 & \mathcal{Q}_2 & \mathcal{Q}_3 & 0 \\
 0 & \mathcal{Q}_2 & \mathcal{Q}_4 & 0 & \mathcal{Q}_4 & 0 & 0 & \mathcal{Q}_4 & \mathcal{Q}_4 & 0 & 0 & \mathcal{Q}_4 & 0 & \mathcal{Q}_4 & \mathcal{Q}_2 & 0 \\
 \mathcal{Q}_2 & 0 & 0 & \mathcal{Q}_4 & 0 & \mathcal{Q}_4 & \mathcal{Q}_4 & 0 & 0 & \mathcal{Q}_4 & \mathcal{Q}_4 & 0 & \mathcal{Q}_4 & 0 & 0 & \mathcal{Q}_2 \\
 0 & \mathcal{Q}_2 & \mathcal{Q}_4 & 0 & \mathcal{Q}_4 & 0 & 0 & \mathcal{Q}_4 & \mathcal{Q}_4 & 0 & 0 & \mathcal{Q}_4 & 0 & \mathcal{Q}_4 & \mathcal{Q}_2 & 0 \\
 \mathcal{Q}_2 & 0 & 0 & \mathcal{Q}_4 & 0 & \mathcal{Q}_4 & \mathcal{Q}_4 & 0 & 0 & \mathcal{Q}_4 & \mathcal{Q}_4 & 0 & \mathcal{Q}_4 & 0 & 0 & \mathcal{Q}_2 \\
 \mathcal{Q}_2 & 0 & 0 & \mathcal{Q}_4 & 0 & \mathcal{Q}_4 & \mathcal{Q}_4 & 0 & 0 & \mathcal{Q}_4 & \mathcal{Q}_4 & 0 & \mathcal{Q}_4 & 0 & 0 & \mathcal{Q}_2 \\
 0 & \mathcal{Q}_2 & \mathcal{Q}_4 & 0 & \mathcal{Q}_4 & 0 & 0 & \mathcal{Q}_4 & \mathcal{Q}_4 & 0 & 0 & \mathcal{Q}_4 & 0 & \mathcal{Q}_4 & \mathcal{Q}_2 & 0 \\
 0 & \mathcal{Q}_2 & \mathcal{Q}_4 & 0 & \mathcal{Q}_4 & 0 & 0 & \mathcal{Q}_4 & \mathcal{Q}_4 & 0 & 0 & \mathcal{Q}_4 & 0 & \mathcal{Q}_4 & \mathcal{Q}_2 & 0 \\
 \mathcal{Q}_2 & 0 & 0 & \mathcal{Q}_4 & 0 & \mathcal{Q}_4 & \mathcal{Q}_4 & 0 & 0 & \mathcal{Q}_4 & \mathcal{Q}_4 & 0 & \mathcal{Q}_4 & 0 & 0 & \mathcal{Q}_2 \\
 \mathcal{Q}_2 & 0 & 0 & \mathcal{Q}_4 & 0 & \mathcal{Q}_4 & \mathcal{Q}_4 & 0 & 0 & \mathcal{Q}_4 & \mathcal{Q}_4 & 0 & \mathcal{Q}_4 & 0 & 0 & \mathcal{Q}_2 \\
 0 & \mathcal{Q}_2 & \mathcal{Q}_4 & 0 & \mathcal{Q}_4 & 0 & 0 & \mathcal{Q}_4 & \mathcal{Q}_4 & 0 & 0 & \mathcal{Q}_4 & 0 & \mathcal{Q}_4 & \mathcal{Q}_2 & 0 \\
 \mathcal{Q}_2 & 0 & 0 & \mathcal{Q}_4 & 0 & \mathcal{Q}_4 & \mathcal{Q}_4 & 0 & 0 & \mathcal{Q}_4 & \mathcal{Q}_4 & 0 & \mathcal{Q}_4 & 0 & 0 & \mathcal{Q}_2 \\
 0 & \mathcal{Q}_2 & \mathcal{Q}_4 & 0 & \mathcal{Q}_4 & 0 & 0 & \mathcal{Q}_4 & \mathcal{Q}_4 & 0 & 0 & \mathcal{Q}_4 & 0 & \mathcal{Q}_4 & \mathcal{Q}_2 & 0 \\
 0 & \mathcal{Q}_3 & \mathcal{Q}_2 & 0 & \mathcal{Q}_2 & 0 & 0 & \mathcal{Q}_2 & \mathcal{Q}_2 & 0 & 0 & \mathcal{Q}_2 & 0 & \mathcal{Q}_2 & \mathcal{Q}_3 & 0 \\
 \mathcal{Q}_1 & 0 & 0 & \mathcal{Q}_2 & 0 & \mathcal{Q}_2 & \mathcal{Q}_2 & 0 & 0 & \mathcal{Q}_2 & \mathcal{Q}_2 & 0 & \mathcal{Q}_2 & 0 & 0 & \mathcal{Q}_1
\end{array}
\right]
\end{align}
where \begin{align*}
\mathcal{Q}_1=&\frac{1}{32} \left(5+\lambda_1\right), & \mathcal{Q}_2=&\frac{1}{32} \left(-1+\lambda_2\right),\\
\mathcal{Q}_3=&\frac{1}{32} \left(5-\lambda_3\right), & \mathcal{Q}_4=&\frac{1}{16} \lambda_4\sinh[\beta ].
\end{align*}
Finally, we can obtain the final density matrix for the TLCQ configuration has the form as:
\begin{align}
\rho_{ISE}(\tau)=\left[
\begin{array}{cccccccccccccccc}
 \mathcal{R}_1 & 0 & 0 & 0 & 0 & 0 & 0 & 0 & 0 & 0 & 0 & 0 & 0 & 0 & 0 & \mathcal{R}_1 \\
 0 & \mathcal{R}_2 & 0 & 0 & 0 & 0 & 0 & 0 & 0 & 0 & 0 & 0 & 0 & 0 & \mathcal{R}_2 & 0 \\
 0 & 0 & \mathcal{R}_2 & 0 & 0 & 0 & 0 & 0 & 0 & 0 & 0 & 0 & 0 & \mathcal{R}_2 & 0 & 0 \\
 0 & 0 & 0 & \mathcal{R}_3 & 0 & 0 & 0 & 0 & 0 & 0 & 0 & 0 & \mathcal{R}_3 & 0 & 0 & 0 \\
 0 & 0 & 0 & 0 & \mathcal{R}_2 & 0 & 0 & 0 & 0 & 0 & 0 & \mathcal{R}_2 & 0 & 0 & 0 & 0 \\
 0 & 0 & 0 & 0 & 0 & \mathcal{R}_3 & 0 & 0 & 0 & 0 & \mathcal{R}_3 & 0 & 0 & 0 & 0 & 0 \\
 0 & 0 & 0 & 0 & 0 & 0 & \mathcal{R}_3 & 0 & 0 & \mathcal{R}_3 & 0 & 0 & 0 & 0 & 0 & 0 \\
 0 & 0 & 0 & 0 & 0 & 0 & 0 & \mathcal{R}_2 & \mathcal{R}_2 & 0 & 0 & 0 & 0 & 0 & 0 & 0 \\
 0 & 0 & 0 & 0 & 0 & 0 & 0 & \mathcal{R}_2 & \mathcal{R}_2 & 0 & 0 & 0 & 0 & 0 & 0 & 0 \\
 0 & 0 & 0 & 0 & 0 & 0 & \mathcal{R}_3 & 0 & 0 & \mathcal{R}_3 & 0 & 0 & 0 & 0 & 0 & 0 \\
 0 & 0 & 0 & 0 & 0 & \mathcal{R}_3 & 0 & 0 & 0 & 0 & \mathcal{R}_3 & 0 & 0 & 0 & 0 & 0 \\
 0 & 0 & 0 & 0 & \mathcal{R}_2 & 0 & 0 & 0 & 0 & 0 & 0 & \mathcal{R}_2 & 0 & 0 & 0 & 0 \\
 0 & 0 & 0 & \mathcal{R}_3 & 0 & 0 & 0 & 0 & 0 & 0 & 0 & 0 & \mathcal{R}_3 & 0 & 0 & 0 \\
 0 & 0 & \mathcal{R}_2 & 0 & 0 & 0 & 0 & 0 & 0 & 0 & 0 & 0 & 0 & \mathcal{R}_2 & 0 & 0 \\
 0 & \mathcal{R}_2 & 0 & 0 & 0 & 0 & 0 & 0 & 0 & 0 & 0 & 0 & 0 & 0 & \mathcal{R}_2 & 0 \\
 \mathcal{R}_1 & 0 & 0 & 0 & 0 & 0 & 0 & 0 & 0 & 0 & 0 & 0 & 0 & 0 & 0 & \mathcal{R}_1
\end{array}
\right]
\end{align}
where
\begin{align*}
\mathcal{R}_1=&\frac{1}{16} \varsigma_1 \left(1+ \varsigma_2 \right), & \mathcal{R}_2=&\frac{1}{16} \left(1-\varsigma_3 \right),\\
\mathcal{R}_3=&\frac{1}{16} \varsigma_4 \left(-1+\varsigma_5 \right)^2.\\
\end{align*}

\end{document}